\documentclass{IEEEcsmag}

\usepackage[colorlinks,urlcolor=blue,linkcolor=blue,citecolor=blue]{hyperref}

\usepackage{upmath}
\usepackage{tabularx} 

\jvol{XX}
\jnum{XX}
\paper{8}
\jmonth{May/June}
\jname{IT Professional}
\pubyear{2021}

\begin{document}

\sptitle{DEPARTMENT: HEAD}

\title{Towards \textit{softer}ware: Enabling personalization of interactive data representations for users with disabilities}

\author{Frank Elavsky}
\affil{Carnegie Mellon University, Pittsburgh, PA, 15213, USA}

\author{Marita Vindedal}
\affil{Highsoft, Vik i Sogn, 6893, Norway}

\author{Ted Gies}
\affil{Elsevier Labs, Miamisburg, OH, 45342, USA}

\author{Patrick Carrington}
\affil{Carnegie Mellon University, Pittsburgh, PA, 15213, USA}

\author{Dominik Moritz}
\affil{Carnegie Mellon University, Pittsburgh, PA, 15213, USA}

\author{{\O}ystein Moseng}
\affil{Highsoft, Vik i Sogn, 6893, Norway}

\markboth{DEPARTMENT HEAD}{DEPARTMENT HEAD}

\begin{abstract}
\looseness-1Accessible design for some may still produce barriers for others. This tension, called access friction, creates challenges for both designers and end-users with disabilities. To address this, we present the concept of softerware, a system design approach that provides end users with agency to meaningfully customize and adapt interfaces to their needs. To apply softerware to visualization, we assembled 195 data visualization customization options centered on the barriers we expect users with disabilities will experience. We built a prototype that applies a subset of these options and interviewed practitioners for feedback. Lastly, we conducted a design probe study with blind and low vision accessibility professionals to learn more about their challenges and visions for softerware. We observed access frictions between our participant's designs and they expressed that for softerware's success, current and future systems must be designed with accessible defaults, interoperability, persistence, and respect for a user's perceived effort-to-outcome ratio.
\end{abstract}

\maketitle

\chapteri{T}here is a significant and relatively unacknowledged problem in emerging work on accessible data representations: a single design cannot satisfy all users. People with disabilities, even those who share the same category of disability, often have different experiences, capabilities, and needs. As experienced practitioners and researchers who have been working to make data representations more accessible (some of us for more than a decade), we have each observed this persistent problem in our own practice.

\begin{figure*}
\centerline{\includegraphics[width=38.5pc]{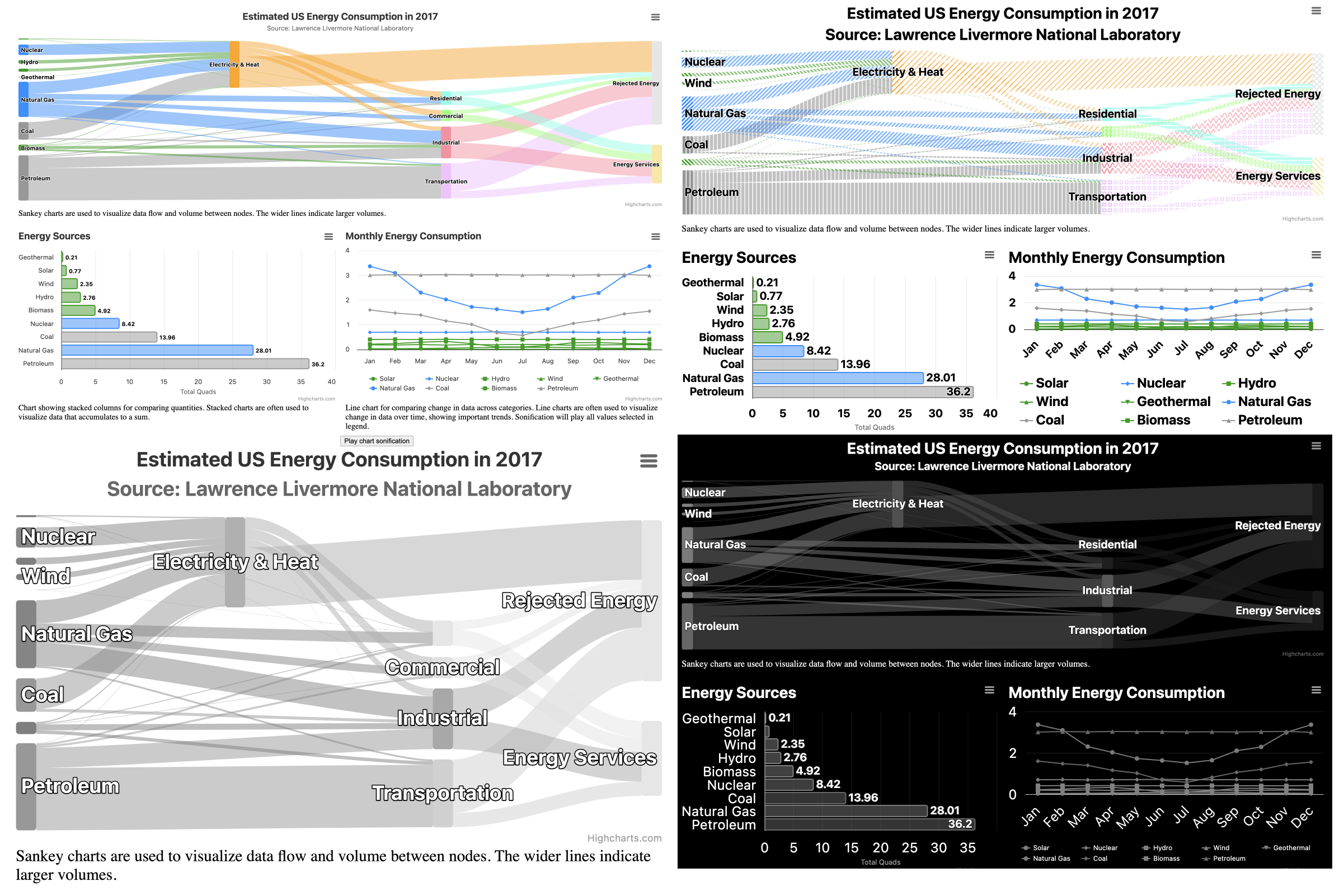}}
\caption{Sometimes one design is not enough. Our design (upper left) and three different designs by low vision users. All low vision users chose larger text, but then diverged: redundant-encoding enabled (upper right), high zoom and greyscale on white (bottom left), and then dark mode (enabled externally) with greyscale (bottom right).}
\end{figure*}

In this work, we focus on data \textit{visualizations} because they are the most prevalent form of data representation in practice. Data visualizations that are produced by a designer for an audience tend to be designed in a way that is relatively \textit{unchangeable}. As a material, we use the metaphor that the creator of a visualization manipulated their design while it was in a \textit{softer} state, like clay. And eventually, the clay is \textit{hardened} into a state that is presented to the user. Often visualization design artifacts cannot easily be altered by an end-user after they are created. Pixels cannot be moved, graphics cannot be re-embedded. The clay has been fired and the visualization is now \textit{baked}. We intend to make the visual and non-visual characteristics of accessible data representations easier to change and manipulate for end-users. We want a material that is softer than software. And while malleable interfaces and end-user programming are a promising starting point for our work, their design principles are often too fluid and unconstrained. We propose to call the space of design between too-hard and too-fluid software system design ``softerware.''

% We wanted to explore material softness \textit{with constraints}. Recent advancements in malleable interfaces and end-user programming both offer ways to enable end-user agency and serve as a useful starting point for our work. However, end-user programming puts a burden on end-users to know the language and symbols of interface-building in order to build their own interfaces and malleable interfaces can potentially lead users into manipulations that are unhelpful or even harmful for their goals and interests. Instead, we chose to enable end-users to have agency over a data visualization interface by exposing a preferences-driven menu of options built from our existing knowledge of visualization and accessibility. We wanted to design a malleable system that was tailor-built to help users overcome accessibility barriers, specifically.

\vspace{4em}

Exploring the \textit{softerware} gap in visualization became our primary focus, which led us to formulate the following qualitative exploratory research questions:
\begin{itemize}
\item \textbf{R1}: What constraints and capabilities should we provide end-users to give them meaningful agency over interactive data visualizations?
\item \textbf{R2}: What qualities, challenges, and design opportunities do designers and engineers envision for a data visualization \textit{softerware} system?
\item \textbf{R3}: What qualities, challenges, and design opportunities do blind and low vision users envision for a data visualization \textit{softerware} system?
\end{itemize}

We contribute our findings from this research to the larger accessibility and visualization communities in hopes that we can inform and inspire future work that investigates \textit{softerware} systems, end-user design, preferences-based user experiences, and malleable interfaces focused on end-users with disabilities. Our future work is dedicated to improving upon and deploying our prototype at scale within Highsoft's Highcharts.

\section{RELATED WORK}
Our contribution is an attempt to bridge the gap between the knowledge we have on accessibility for visualization (as a complex space of design and engineering) with research and practice that centers on users with disabilities being able to adust, change, or control the interfaces they interact with. We intend to frame our work towards the benefit of data visualization designers, system engineers, and end-users of data visualizations. We believe that more flexible data visualization systems that enable user preferences will require a careful approach to architecture and thorough consideration for the burdens placed on end-users.

\subsection{Data Visualization and Accessibility}
Data visualization accessibility has come far in recent years. But little work has been done to explore what disability scholars call ``access friction'' - a tension that arises when access must be negotiated \cite{Hsueh}. This friction is often a result of static barriers in shared spaces: one artifact or approach designed to include some people may end up excluding others.

In general, accessibility concerns itself with a broad spectrum of barriers that people with different disabilities face. And while most literature focuses on visual disabilities \cite{Wimer}, there are growing resources on areas such as cognitive/intellectual disabilities \cite{Wu}, neurodivergence \cite{Tran}, and research exploring epilepsy and vestibular/motion inaccessibility in visualization \cite{South}.

% access for assistive technologies that are used by people with dexterity and upper-body motor impairments \cite{Elavsky2023}, 

Yet despite these resources, making data visualizations more accessible remains a difficult task for practitioners \cite{Joyner}. Some accessibility guidelines even conflict, for example on the topic of patterns and textures used in charts. One side stresses that patterns are harmful to cognitive and visual accessibility while another stresses that redundant encoding strategies are necessary \cite{Elavsky2022}. Understanding how to make the correct design decisions may sometimes be impossible. Either existing guidelines are incorrect or it is possible that access friction becomes inevitable the more we know what barriers exist for different people.

\subsection{Systems that Adapt}

One angle of exploration that has been engaging this issue already focuses on systems that can adapt. Work on adaptive systems for people with disabilities, such as in \textit{ability-based design} \cite{Wobbrock2011}, stresses the importance of design alleviating burdens placed on users. Users who don't fit initial system designs are often expected to adapt to fit the system. This means that they may have to acquire an assistive technology, learn a peripheral skill, hack the system, or wait on a design fix. This places the burden on the user to fit the system. Ability-based design instead stresses that systems should be capable of automatically adapting, in order to reduce these burdens placed on the system's users. 

However, building data visualizations that automatically adapt to users via some form of data collection often do so through means such as monitoring live biometric data and input patterns, collecting a user's self-declared conditions and cognitive ability, parsing a user's history, and sensing a user's environmental or situational context \cite{Yanez}. We argue that these methods for an adaptive system raise questions of end-user agency, trust, privacy, and awareness in regards to the system decision-making. They may not be sufficient for addressing a user's needs while also preserving their privacy and agency.

% \cite{Pallapa}

\subsection{Personalization and Accessibility}
Lastly, we researched broader spaces where users have more design agency and explicit awareness of a system that is built to be adapted. We were interested in literature and projects that explore ways end-users can enact meaningful change on an interface, with special attention paid to accessibility and disability.

One specific project has emerged at the intersection of accessibility, visualization, and customization which focuses on screen reader users adjusting the content of textual tokens when navigating data visualizations \cite{Jones}. While this is excellent work, we still have larger questions about when preferences, options, and customizations are appropriate and in what contexts as well as other ways of conceptualizing end-user agency over a system. It remains unclear when, why, and how customization and personalization can be used effectively when designing a system.

In the field of meta-design, meta-designers consider these end-user manipulations of a system to be one facet of ``end-user design'' and ``continuous co-design'' between a system and a user \cite{Maceli2013}, which helps give us some meaningful language to refer to our system goals.

Recent work on the influential factors for personalization and adoption of accessibility settings \cite{Wood2023} also informs our work in 2 key ways: conceptual mismatching between a system and user can contribute to a system's under-use while features that propose value, are time-saving, or reduce cognitive load for a user can contribute to positive perception and use of personalization of a system.

\section{PRESENTING: \textit{SOFTERWARE}}

\textit{Softerware} is a vision for software design that is not just based on giving a user the ability to set preferences or personalize. \textit{Softerware} is about the intentional design of a software system that enables people with disabilities to have meaningful, opinionated, and persistent agency over that system.

While a venn diagram of \textit{softerware}, end-user programming, and preferences-based parameter controls would have significant overlaps, \textit{softerware} has distinct differences. Both end-user programming and preferences-based parameter controls inform our work.

However, \textit{softerware} differs from end-user programming by prioritizing accessibility-centered constraints and language that align with users' lived experiences, rather than exposing general-purpose interface-building constructs focused on system- or task-centric language. It can be seen, largely, as a subset of some flavors of end-user programming.

Additionally, \textit{softerware} can be differentiated from common preferences-based parameter controls because it does not simply expose a set of configurable options, but instead frames these options through anticipated access barriers, prioritizing their discoverability, relevance, and impact on usability. Also, notably, \textit{softerware} is concerned with the design of systems, where parameter-based solutions are just one possible design approach. We imagine that \textit{softerware} also includes broader system design directions than preferences-based parameter controls.

We contribute the concept of \textit{softerware} to the larger community of researchers and practitioners because we argue it is a useful construct that can help us categorize past work, improve existing projects, and inspire new directions. \textit{Softerware} systems have been part of existing work for decades, but we lack a cohesive way to refer to designing and engineering experiences that enable end-users to have agency over interfaces, especially in the field of data visualization.

\subsection{Defining \textit{Softerware}'s Principles}
Here we present the principles that define \textit{softerware} before demonstrating an example instantiation in the context of online, interactive data visualization.

\subsubsection{Principle: Has Reasoned, User-centered Constraints}
An important aspect of \textit{softerware} is that it is \textit{softer} than software (which is already-baked) but not quite as \textit{malleable}, free, and potentially low-level as systems that facilitate fully realized end-user programming \cite{Cabitza}.

End-user programming is still a form of \textit{programming}. It centers on constructs, functionalities, and reasoning from software programming and translates these elements to users in ways that may suit a user's natural language or mental models, such as through no-code, visual-only, or low-code approaches. However, we anticipate that users experiencing accessibility barriers will prefer not to interact with \textit{any} language that centers on software paradigms.

Instead \textit{softerware} engages this limitation through reasoned constraints that leverage conceptualizations and language focused on overcoming anticipated user barriers. \textit{Softerware} provides constraints and then frames and presents those constraints in ways that have vocabulary correspondence to user needs.

To accomplish this, the \textit{softerware} system designer must work to anticipate not only what their system should do in a default or beginning state but also which ways that system will potentially fall short and require fitting by the end-user. The system designer should motivate all of the capabilities of a \textit{softerware} system based on what they anticipate users will want to change, how users can discover that change is possible, and then how best to enable users to enact that change easily.

\subsubsection{Principle: Facilitates End-user Agency}
\textit{Softerware} is ultimately about the process of architecting and implementing a system that enables an end-user to be able to easily express meaningful changes to that system's appearance and behavior.

Accessibility has been framed as a tension between fit and scale \cite{Hickman}, where \textit{fit} refers to a system that is perfectly complimentary and synchronized to a user and \textit{scale} refers to a system that is capable of reproducing functionality for many different users. We believe that the tension between fit and scale, in addition to \textit{access friction}, can both be alleviated when a system is designed to facilitate end-user agency.

The cornerstone goal of a \textit{softerware} system is an attempt to facilitate \textit{self-fitting} at a minimum, and in ideal circumstances also facilitate social methods of sharing fitting (such as loading profiles or ingesting metadata from others).

\subsubsection{Principle: Demonstrates Value}
Existing literature makes one thing particularly clear when it comes to personalization and end-user design: it has to be worth it \cite{Wood2023}. Users must be able to recognize barriers, issues, or shortcomings of a system and then discover and utilize capabilities provided to them to eliminate or alleviate those barriers.

This entire process must not be too burdensome and the payoff should establish an expectation that future use of the system will be improved. The time and effort it takes for a user to fix a problem should be less than the time and effort generated by that problem. This means that \textit{softerware} systems can likely be optimized and improved significantly over time, as better techniques are developed to perform tasks as quickly and easily as possible.

The user should also be able to validate the value of their interaction with a \textit{softerware} system through the continued use of that system. If something was a painful experience and they took action to alleviate that issue, they should be able to observe the effects easily.

\section{PROTOTYPE: VISUALIZATION \textit{SOFTERWARE}}
\begin{figure*}
\centerline{\includegraphics[width=38.5pc]{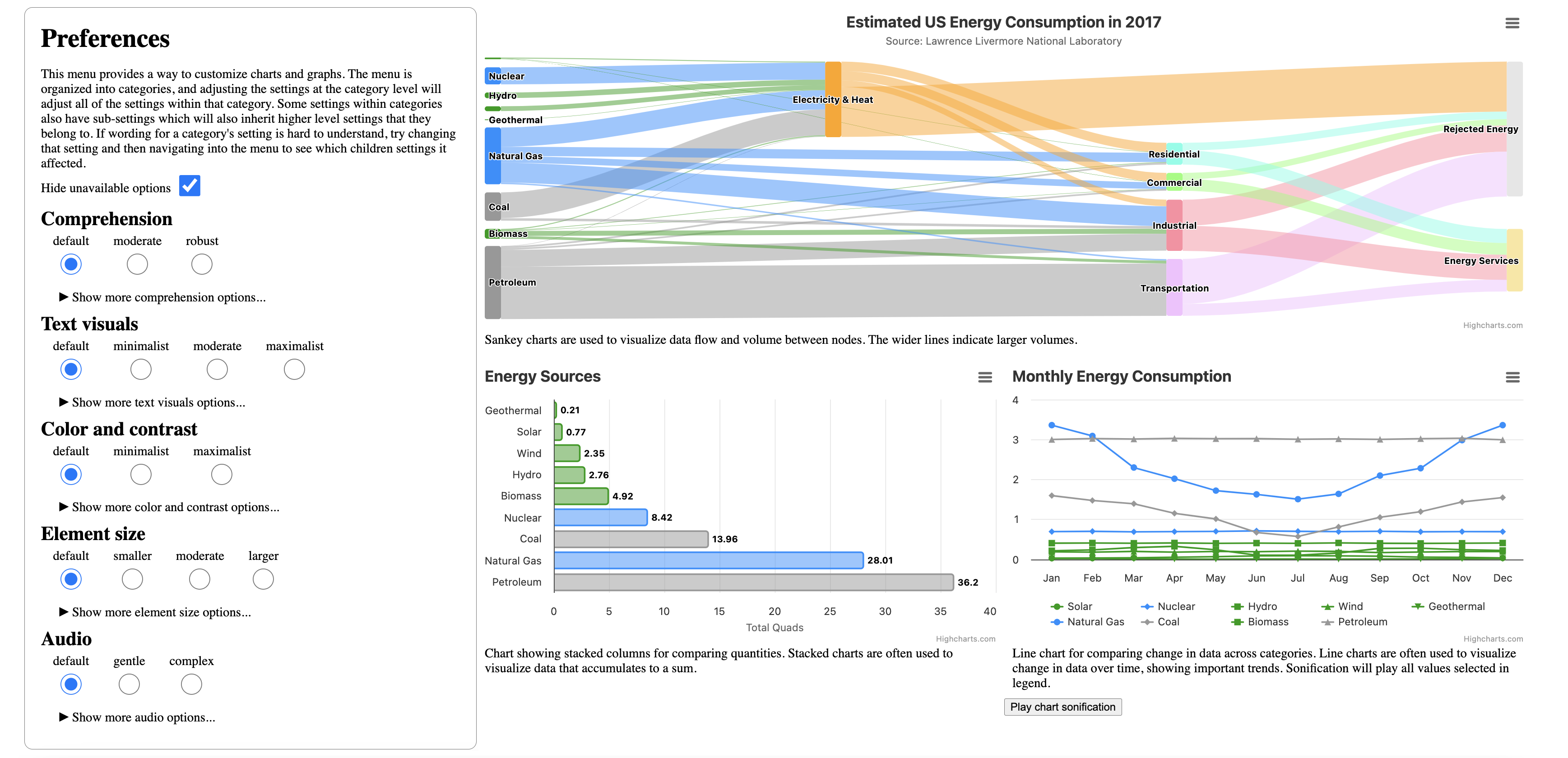}}
\caption{Our dashboard design on US Energy Consumption in 2017 with a sankey, bar chart, line chart, and preferences menu on the left.}
\end{figure*}

We first built a data visualization dashboard (see {\bf Figure 2}) that would allow us to build a \textit{softerware} system prototype that we could demonstrate to designers, engineers, and evaluate with end-users with disabilities. You can view and interact with our prototype, including view our \href{https://github.com/highcharts/highcharts-a11y-prototyping}{open source code and dataset on user preference options, on github}.

We chose a relatively clean and standard dataset that would be relevant to our target end-users, who we would recruit from the United States, on 2017 US energy consumption. This dataset afforded us enough complexity to build multiple data visualizations in one dashboard (including an uncommon type like the sankey). This choice also allowed us to explore more ideas for user interventions and investigate broader questions in our eventual study. Our dashboard was built in JavaScript and laid out in a wide format for interacting with on a desktop machine.

We designed the dashboard to contain some interactivity, but nothing highly complex. Each chart has tooltips and visual filtering provided on hover/keyboard focus and the line chart has data filtering through the legend as well as sonification.

\subsection{\textit{Pretty Accessible} by Default}
In order to really test \textit{access frictions}, we wanted our dashboard to be considered accessible in its default state. We ran manual and automated tests according to accessibility standards as well as a 50-heuristic manual accessibility evaluation specific to interactive data visualizations \cite{Elavsky2022, WCAG}. In addition, we chose to use Highcharts to create the visualizations in our dashboard because existing work has demonstrated the breadth of their accessibility capabilities \cite{Kim2023}.

We ensured that text contrast was strong, text size was well above guidelines, interaction targets were of a minimum size, screen reader access was descriptive and interactive, our DOM was structured into a hierarchy, we provided semantic HTML data tables for each chart, and there were no \textit{critical} issues detected when running automated tests using \textit{axe DevTools} software. Many of the capabilities that made our dashboard accessible were provided out of the box by Highcharts.

\subsection{Reasoned Constraints: 195 Accessibility Options for Interactive Data Representations}

After building our initial dashboard (see Figure 2), our research team collaborated and discussed potential access frictions that might arise from the use of our dashboard. We used our existing experience from accessibility work, more than a decade in the case of the Highsoft and Elsevier co-authors, to assemble a list of concrete, expected user barriers that would be hard to resolve in a single design, and organized those barriers based on common themes.

We then used these themes to brainstorm how we anticipated end-users would identify barriers and then what language they might use to describe an identified barrier and overcome it. One example might be under the theme ``hard to see'': ``I can't read this text'' as a barrier with the final language for them overcoming that barrier as an expression similar to, ``I wish this text size was larger'' or ``make this text bigger.''

From these solutions, we re-framed the language into interactive option categories. All categories were given binned options and not more than 7 choices, to avoid overwhelming users. For example, ``text size'' was an option category and had 7 binned choices from ``small'' to ``large'' available. We then created a hierarchy of option categories underneath these higher-level categories which could allow for more element-level specificity within a data visualization.

% So ``text size'' as a higher level category with options also contained children that had more specificity, such as ``title text size,'' ``legend text size'' and so on.

The final stage of our language and options design was in line with existing work on use of preference-based interfaces, which is to avoid organizing the language of our configurations based on categories of disability \cite{Bailey} and instead focus on higher-level categories of identifiable elements of a user's experience, such as ``text visuals,'' ``audio,'' and ``color and contrast.''

At the end, we produced 195 option categories and 774 total option choices. Using the combinatorial rule of product, we calculated 6.83e114 possible unique end-user design configurations, which is more than the estimated number of atoms in the universe. 

\subsubsection{Narrowing the scope of our personalization options}

Despite the breadth of this initial exploration of categories and options, a system that has 195 categories and 774 options would put too much of a burden on the cognitive load of our participants. We believe that even a minimalist interface design that surfaces all of these choices to users would conflict with our principle to \textbf{demonstrate value} for a user's time and effort.

So, to ensure both that scope of our user study was feasible and that we maximize the value for our participants, we reduced our initial working categories down to 33 with 137 total options, all focused on options we believed would be most relevant specifically for users who are blind or low vision. We collaborated as a group reach consensus on this subset of categories and options which we filtered according to the following criteria. To guide our work we first only considered options that are explicitly referenced in web standards, research, or guidelines for accessibility as it relates to visualizations~\cite{Elavsky2022} as such: (1) Options that influence alternative text customization, (2) contrast and redundant encoding, (3) text readability, (4) high level control of sonifications, and (5) that would be feasible to implement in Highcharts (some of our categories, specifically those related to styles of interactivity, would require design iterations).

These options and our subset are viewable in \href{https://highcharts.github.io/highcharts-a11y-prototyping/early_explo/examples/menu/menu.html}{our live, interactive demo} online.

\subsection{Preferences Menu Design}
We then iterated on visual and functional designs to allow users to actually interact with and enact these design configurations. Our early ideas included a natural language interface (since we used a relatively ``natural language'' centered process to develop these categories), direct manipulation of the elements in a visualization (through focus, hover, click, or selection methods), and eventually settled on the user interface of a separate, visually nearby menu with nested options (see Figure 2). We designed our menu so that manipulating higher level options in the hierarchy would enact downstream options to follow suit, but any manipulation to downstream options would override higher controls, following common patterns used in systems that implement hierarchical specificity.

We justified our user interface as a menu for our final choice because it provides a place for metadata from the other design ideas (natural language and direct manipulation) to live, in case we develop those down the line as well. We anticipated that a menu not only provides a means of interaction but also storage of the state of a system. In addition, this type of user interface is common and relatively recognizable.

On the last point, we argue that using a familiar, simple, and well-established design as our starting point enables our research to explore more foundational questions. We anticipate that designs focused on direct manipulation, natural language (chatting), or even automatically adapting would introduce additional cognitive burden for our users, which may complicate our study design and influence our findings.

\section{EVALUATION}
Our first research question for this project (``What constraints and capabilities should we provide end-users to give them meaningful agency over interactive data visualizations?'') focused on our thematic collation and compilation of anticipated access frictions, but our following two research questions would require outside evaluation: ``What qualities, challenges, and design opportunities do designers and engineers envision for a data visualization \textit{softerware} system?'' and ``What qualities, challenges, and design opportunities do blind and low vision users envision for a data visualization \textit{softerware} system?''

\subsection{Preliminary: Visualization Practitioners}
The preliminary step in our evaluation was to investigate what qualities, challenges, and design opportunities data visualization engineers and designers envision for a \textit{softerware} system.

\subsubsection{Recruitment}
We recruited 4 data visualization practitioners, each with roles as a current or former visualization software engineer (3) or designer (1). We recruited participants from our existing network of engineers and designers, requesting participation via email. Our practitioners were not compensated for their participation and we asked them up front if they would be willing to volunteer their time for us.

\subsubsection{Procedure}
We conducted 30-minute, semi-structured, qualitative interview sessions either over Zoom or in-person. The session consisted of a 5 minute explanation, 5 minute demo of our prototype's capabilities, and a series of open-ended, semi-structured questions for 20 minutes. Our questions started with getting their thoughts on the idea, what they anticipated other developers and designers would think, what aspects of a visualization they believe end-users will want control over, issues they believed end-users would face, and what new opportunities they envision our prototype and underlying design concept of \textit{softerware} enables.

\subsection{Study: Blind and Low Vision Users with Accessibility Expertise}

Our primary study was focused on our third research question on the qualities, challenges, and opportunities that users with disabilities, in this case users who are blind and low vision, envision for a \textit{softerware} experience of interactive data visualizations. To explore this, we used our prototype dashboard as a design probe to stimulate concrete feedback and ideation on both the details of our prototype as well as our larger design concept of \textit{softerware}.

Our study is intended to contribute qualitative knowledge, largely because we believe that statistical generalizations or controlled experiments about a particular group or subgroup of people with disabilities may not lead us to explore \textit{access friction}. We want to explore ways that broader guidelines and design knowledge are capable of producing artifacts that still retain barriers for some individuals with disabilities. This larger challenge (that general guidelines may not provide a meaningful fit for all individuals with disabilities) is not new to accessibility research \cite{Power}.

And to this end, we designed our study to maximize the production of knowledge that is considerate and careful of individual differences, challenges, preferences, and envisioned opportunities.

\subsubsection{Recruitment}
\begin{table}
\caption{Study Participants}
\label{tab:participants}
\tablefont
\begin{tabular*}{18pc}{@{}llll@{}}
\toprule
PID & Age & Gender & Disability \\%                                   & Tech                \\
\colrule
P1  & 39  & F      & Totally Blind \\%                                & NVDA, Windows       \\
P2  & 38  & F      & Totally Blind \\%, Left Arm Motor Impairment \\%     & JAWS, Windows       \\
P3  & 46  & M      & Legally Blind \\%Blind w/ Partial Light Perception \\%            & ORCA, Linux         \\
P4  & 28  & F      & Low Vision \\%, Light Sens., Nightblindness \\%      & Magnification, Mac  \\
P5  & 34  & M      & Legally Blind \\% & Various, Windows    \\
P6  & 52  & M      & Totally Blind \\%                                & JAWS, Windows       \\
P7  & 56  & F      & Low Vision \\%, Glaucoma (both) \\%                  & NVDA + Mag, Windows \\
P8  & 36  & M      & Low Vision \\%, Cataracts (both) \\%                 & Various, Windows   
P9  & 55  & M      & Totally Blind \\%, Cataracts (both) \\%                 & Various, Windows   
\end{tabular*}
\end{table}
Our study involved 9 total participants who are blind or low vision (see \textbf{Table 1}), all of whom are also professionals with accessibility expertise (either currently or formerly employed in an accessibility-specific role as subject matter experts). 5 of our participants self-identified as male, 4 as female. Average age of our participant group was 42.67 (SD = 9.99). We initially recruited 6 participants using an existing, compensated research relationship between Highsoft and an external consultancy. In addition, we recruited 3 more participants from our existing network of accessibility consultants, who were each compensated 100 USD for their time.

We anticipated that recruiting participants who not only have lived experience with a disability but also are subject matter experts in accessibility would contribute to the depth of our qualitative study as well as general breadth of considerations. We wanted to maximize the value of feedback on our work.

We reached out to all participants via email with a call for participation and participants were screened according to whether they are blind or low vision. Participants were notified in advance of compensation and that consent to participate is voluntary.

\subsection{Procedure}
Our qualitative study sessions were recorded and conducted over zoom in 3 primary phases (plus a break) during one 90 minute session. Our phases were: early interview, task-evaluation of our dashboard (menu hidden) with discussion, a break, and task-evaluation of our dashboard (menu shown) with final discussion.

We structured our study with 2 data tasks, one elementary and one synoptic \cite{Aigner}, for their exploration of our prototype with and without the menu. All 4 tasks differed between the two versions of our prototype.

Our intention for asking participants to perform tasks was not to measure their speed or accuracy (or other objective measures), but simply as a probe for eliciting feedback on the usability and effectiveness of our prototype and design before and after introducing our menu prototype. The tasks served as a way for us to situate and focus our semi-structured interview in a simulated use of a dashboard environment and on the differences in qualities the participants experienced with and without the ability to personalize.

\subsubsection{Introduction and Early Interview [20min]}
Our session opened with an introduction to the research team and gathering verbal consent from participants for participation. We gathered demographic information from participants and asked them about their current assistive technology use. We followed up with questions related to whether or not they customized their technology in any way, through adjusting settings, modifications, adding scripts, getting extensions, or equivalent. We then ended the opening session with ice-breaker questions about whether they can recall a chart or graph they have experienced in the past and what their favorite way to experience a chart is.

\subsubsection{No Menu Prototype, Tasks, Discussion [30-35min]}
The next phase of our session involved showing participants our demo environment (see Figure 2), except that our preferences menu was hidden. We explained what the dashboard was, including explaining each chart type shown and how to read them. We gave users a short amount of time to explore the dashboard, and then notified them that in order to evaluate the effectiveness of our technology, we would be asking them to perform 2 data tasks.

Our first data question was for participants to perform an elementary analytical lookup task, ``Does petroleum or nuclear contribute the most to Electricity and Heat?'' (``nuclear'' was correct). Our second task was, ``Which energy type has the highest use in Dec \textit{and} Jan?'' (``natural gas'' was correct). We gave participants a limited time (5mins total) to answer the questions and upon answering, we gave them the correct answer and asked them to explain their process of finding their answer, step-by-step.

Our final step in this process was to interview them about their perceived challenges and frustrations with the dashboard and whether anything could be changed or adjusted in order to help them complete their tasks. We followed this phase with a 10 minute break.

\subsubsection{Menu Prototype, Tasks, Discussion [30-35min]}
We opened the final phase of our study by sending our participants a new link to a version of our online dashboard that included our preferences menu. We explained the purpose of the menu and gave them 5 minutes to explore the available options. While this was a relatively small period of time to explore the available options, due to our time constraints we did no want users to perform data tasks 

After participants explored the menu and its effects, we repeated our tasks procedure. Participants were given 2 tasks to complete in five minutes. First, an elementary analytical task, ``Where does most coal go?'' (``electricity and heat'') and then a synoptic task ``In the summer, June through August, which energy type has the highest consumption rate?'' (``petroleum'').

Our final discussion focused on investigating our participant's thoughts on our prototype, the idea of preferences and customization, why they chose the customizations that they did, whether they had any new or additional ideas, considerations for other users with disabilities, and any other concerns, challenges, or feedback. We asked them specifically to consider both their personal, lived experience with their disability and assistive technology in addition to their professional expertise in accessibility.
% \begin{table}
% \caption{Technology Use}
% \label{tab:use}
% \tablefont
% \begin{tabular*}{18pc}{@{}lll@{}}
% \toprule
% PID & Technology & Custom? \\%                                   & Tech                \\
% \colrule
% P1 & NVDA, Windows & Yes      \\
% P2  & JAWS, Windows & Yes      \\
% P3  & ORCA, Linux  & Yes       \\
% P4  & Magnification, Mac & Yes \\
% P5  & Magn., Browser Ext., Windows & Yes   \\
% P6  & JAWS, Windows & Yes      \\
% P7  & NVDA, Magn., Windows & Yes \\
% P8  & Magn., Sys. Contrast, Windows & Yes  \\
% P9  & JAWS, Windows & No 
% \end{tabular*}
% \end{table}

\section{RESULTS}

We performed two analyses from our studies, first analyzing our findings from our preliminary study from practitioners and then analyzing our results from our study with end-users. For our analyses, we collated our notes and transcript materials, coded them thematically, and then used affinity diagramming to group the themes that emerged from our data. For instances where our principles are illustrated in our results, they will be called out \textbf{in bold}.

\subsection{Preliminary Findings}
To avoid repeating information between our preliminary study with visualization practitioners and final study with blind and low vision participants, any findings from our end-users that are echoed by our practitioners will be mentioned later. Only the findings unique to our preliminary study will be included here.

\subsubsection{Alleviating Situational Barriers}
3 of our 4 practitioners spoke about the potential benefits of \textbf{end-user agency} of a visualization for situational or contextual reasons. One participant gave the example that when giving a presentation using an existing dashboard, having the ability to manipulate features to suit layout, flow, and interactions on-demand would be valuable. Another example given was that at times a user's viewing device (such as a smartphone) can cause barriers, so softerware would be useful to have available.

% , mirroring sentiments from existing literature on the benefits of accessible technology for situational barriers faced by people who are not disabled \cite{Wobbrock2019}

\subsubsection{Creating Potentially Harmful Visualizations}
The second theme from our practitioners (3) was the concern that end-users would be able to create a misleading or harmful data visualization. For example, encoding area size and aspect ratios can both be misleading or deceptive, yet being able to manipulate these for accessibility and contextual barriers (such as viewing a chart designed for desktop on a mobile phone) are important design considerations. \textit{Softerware} systems must be designed with \textbf{reasoned constraints} to prevent harmful end-user design.

% Area size sometimes affects users with upper body physical or motor impairments (such as those with tremors) when a chart is interactive with a pointer, while aspect ratios are important for users with low vision being able to adequately find and compare visualizations and see visualizations when using high levels of zoom.

\subsubsection{Designing via \textit{Softerware}-first}
The final theme from our practitioners was around authoring and design-tuning via \textit{softerware}, where 3 participants discussed using a direct-manipulation or LLM-based \textit{softerware} interface to author data visualizations and 1 of the 3 also mentioned that large-scale, privacy-preserving data collection from users could be used to create smarter design defaults in the future. We believe that both suggestions would \textbf{demonstrate more value} to users over the lifetime of a system.

\subsection{Prototype-level Feedback}
Because our participants had accessibility expertise in addition to their lived experience, we were able to get feedback on our existing prototype as well as on our larger idea space for \textit{softerware}.

\subsubsection{Navigation Structure Options}
While not a theme across participants, P2 mentioned that they would like to be able to navigate a data visualization using headings with their screen reader. This was a suggestion that immediately led to our team iterating in parallel on ideas because it provides screen reader users more navigational \textbf{agency}.

% Most screen reader users navigate information via headings when first encountering a new web page \cite{WebAIM}. This suggestion made sense to explore as a sensible default.

\subsubsection{Previewing Change}
Our low vision users (P4, P7, P8) requested a feature to directly show what different options would look like in the preferences menu. For example, the ``text size'' options would display the text size shown for each option in the menu. Low vision users in particular often use high levels of magnification and zoom, so the live results shown in the visualization space required users to go back and forth between the menu once an option was chosen and back into the chart space. In-menu previewing would reduce their overall effort, providing greater \textbf{value}.

\subsubsection{Language Re-consideration}
Some of our participants (P1, P2, P4, P6, P7, P8) noted ambiguity or lack of clarity in the wording we used for our menu's higher level options. For example, ``Sonification order'' under ``Audio'' had the options ``default,'' ``sequential,'' or ``simultaneous.'' ``Gentle'' in ``Audio'' would set the child setting for sonification order to ``sequential,'' but this was unclear initially.

Other participants (P1, P2, P3, P6) noted that while the menu's focus on functional categories was helpful, it might be nice to also have a way to customize the menu itself or view it from a ``disability'' perspective, so they could get all the screen reader options in a single place. Users were interested in looking at all options relevant to ``screen readers'' or ``low vision'' together.

Re-considering our language stresses the importance of \textbf{reasoned, user-centered constraints}, reducing the user's cognitive effort translating between our system and their goals.

\subsection{System-class Accessibility Findings}
The next set of themes that emerged were considerations that both our end-users and our visualization practitioners shared, which we frame as \textit{system-class} considerations \cite{Fleizach}. This theme emerged thanks to the accessibility knowledge and expertise of our end-users and engineering concerns of our practitioners and we believe will be especially helpful for future system work.

\subsubsection{Persistence}
Every participant (P1, P2, P3, P4, P5, P6, P7, P8, P9) as well as all of our practitioners noted that the ability to create some sort of ``profile'' or persistent state of their customizations would be one of the most important features that would make \textit{softerware} actually useful and \textbf{demonstrate value}.

\begin{quote}
``What if I come back to this? Will I lose this? Do I need to do it again?''---P6
\end{quote}

We followed up by asking whether certain contexts would make persistence more or less important. We asked users whether a random website or news article with a chart in it would be worth their time, to which most users replied, ``no.'' However, P5 noted that ``This is so fun that if it was there I still might play around with it and use it, especially if I had the time.'' P4 related this issue to an existing frustration with video games, noting that having to set up options for every game was time consuming. It would be nice if they could ``do this once and forget it.''

\subsubsection{Profile Sharing}
Following this theme of establishing a profile, most of the participants (P1, P3, P4, P5, P6, P8) and 2 of our practitioners also expressed interest in being able to \textit{share} and \textit{import} profiles, in order to save others or themselves time. We believe that enabling social support for \textit{softerware} could improve \textbf{user agency} as well as \textbf{demonstrate more value} to users with different levels of interest in personalization.

In phase 1 of our procedure, we asked users about their existing levels of modification, customization, and preferences setting in their existing use of technology. While all of our participants (except for P9) customize, personalize, or modify their technology to some degree, some (P3, P4, P5, P8) were also the most passionate about being able to save \textit{other} people time and not just themselves.

\begin{quote}
``I customize my tech a lot. If I use something for the first time and it feels off, I find a way to fix it. But most people aren't like that; it takes too long. So I love when I can share [my modifications and customizations] with others.''---P3
\end{quote}

\subsubsection{Cross-system Interoperability}
Closely related to \textit{persistence} and \textit{profile-sharing} was an idea expressed by several participants (P3, P4, P5, P7, P8) that they wanted to be able to use these settings outside of Highcharts. ``Will this work in Microsoft Excel?'' and ``I use Salesforce for analytics a lot and would love this there,'' remarked P4. In order to \textbf{maximize value} for users, it makes sense that they should be able to use their profile with \textit{any} data representation interfaces.

% However, cross-system interoperability would require multiple charting libraries being intelligent enough to ingest user settings, when most are currently incapable of even recognizing a system's ``high contrast'' settings being active.

In addition, all 4 of our practitioners suggested that there would need to be a system in place, either at the operating system level or as some kind of service hosted by a platform, where these settings could be recognized and ingested. For cross-system interoperability to be made possible, it would require establishing standards for customization and preserving user privacy.

\subsection{User-Centered Findings}
The last major set of themes is related to the considerations of end user experience of a \textit{softerware} system applied in practice, including our observations about the differences between users and their choices when personalizing a data visualization interface.

\subsubsection{Frictions in User Differences}
Our first major user-centered finding was that no participant chose the same set of preferences as another. Every user discussed different reasons for justifying their choice of options. Users even chose options that others specifically emphasized were inaccessible to them. An example of this was a tension in preference for and against use of ``dark mode'' designs. P4 mentioned that they had ``night blindness'' (\textit{nyctalopia}), which is why dark mode designs are helpful for them.

\begin{quote}
    ``If anything has dark mode? That's great. I wish everything used dark mode.''---P4
\end{quote}

However, P7 also mentioned that they had progressive nyctalopia, but dark mode makes an interface ``virtually impossible'' to them.

\begin{quote}
    ``Oh, I can't use dark mode at all. I hate when websites have [dark mode] because it can be virtually impossible to use.''---P7
\end{quote}

Any one of the designs chosen by a low vision participant would have been insufficient for providing access for any of our other low vision participants (see Figure 1), highlighting the necessity of \textbf{user agency} over any visual interface design.

Some of our blind participants justified their differing preferences on audio and text accessibility with similar reasoning, for example arguing for cognitive accessibility when deciding on text description length. For example, P9 stressed that ``I prefer to keep things simple'' to ``avoid overwhelm'' while P2 said, ``more information is better than less, when it comes to data.'' Both P9 and P2 preferred ``accessible defaults,'' but disagreed on what length should be default.

\subsubsection{Accessible Defaults are a Necessary Prerequisite}
Several participants were concerned that this approach would allow designers and developers to continue to make inaccessible charts (P2, P6, P7) if users have the ability to \textit{self-fit}. Participants emphasized how important it is to have strong accessibility \textit{before} customization is introduced (P2, P4, P6, P7, P9). Even 3 of our 4 of our practitioners expressed worry that \textit{softerware} could put a design burden on users.

P9, our only participant who almost exclusively uses default settings (and avoids mods and extensions) with their current assistive tech, stressed the importance of well-thought out defaults. It is clear that for users like P9 in particular, strong defaults are much more important than customization. Some assistive technology users are not interested in the work involved in personalization and would prefer technology to suit their needs out of the box.

This leads us to argue that there is a line between ethical use of \textit{softerware}, which is built on top of already-accessible material, and \textit{softerware} that is filling gaps in poor design. Designs that lack fundamental accessibility, such as simply having alt text, fail to \textbf{demonstrate value} in the first place.

\subsubsection{Effort-to-Outcome Ratio}
As a playful rephrasing of the visually-centric (and controversial) \textit{data-to-ink} ratio, we observed an \textit{effort-to-outcome} ratio among our participants, which is fundamentally linked to our principle that \textit{softerware} must \textbf{demonstrate value} to users for the time and energy they put into interaction. Nearly all participants (P1, P2, P3, P4, P6, P7, P8, P9) noted that the work required to interact with this menu wouldn't be worth the effort if they had to do it every time they interacted with a data visualization.

Most of our participants who used screen readers (P1, P2, P3, P6, P9) also mentioned that the menu itself was too cumbersome for navigating within and back and forth with the dashboard. Keeping what a previous state was like in memory was hard, stressing the difficulty designing \textit{softerware} systems that \textbf{maximize value} for a user's effort.

P6 was interested in different ways that this process could become easier,

\begin{quote}
    ``What if I could just tell it what to change while I'm listening? Like right here [navigating a chart element] what if I could just say ``keep it short'' or maybe ``wait, tell me more.''---P6
\end{quote}

This suggests that there may be a space to explore non-visual direct manipulation \textit{softerware} strategies.

\section{LIMITATIONS}
Our work has a limited scope, overall. We intentionally chose evaluate our \textit{softerware} prototype with users who are blind and those who are low vision, in order to elicit access friction and justify the use of our \textit{softerware} solution. However, significantly more work must be done to understand how to design \textit{softerware} systems that support people with other disabilities.

Additionally, as participants in our study suggested, there are other interface design choices that are possible if there is an underlying \textit{softerware} infrastructure available. Language-driven, direct manipulation, and multi-modal interfaces could all be used to explore \textit{softerware}, and our work does not address how to make these interfaces possible or successful.

Lastly, our work focused on a \textit{softerware} prototype for Highcharts. In order for other visualization tools to be built with \textit{softerware} infrastructure, more work will need to explore scalable standards for visualization personalization.

\section{CONCLUSION AND LOOKING FORWARD}
In an idealized world, designers do their best to produce useful and accessible interfaces. They're concerned with making software as accessible as possible by default. But no single design is capable of perfection. \textit{Access frictions} between accessible defaults and the needs of real individuals might always be present in software interfaces. To that aim, we hope to contribute knowledge that can inform future designers and developers to not only build accessible artifacts, but build \textit{systems} that enable end users with disabilities to have interactive agency over their software experiences.

In light of our findings, visualization tool designers are encouraged to reframe their approach to accessible design by incorporating the softerware principles discussed in this work. Tool designers should consider moving beyond static, one-size-fits-all solutions and instead build platforms that support dynamic, user-driven customization. This entails incorporating accessible defaults that serve as a strong foundation while also providing intuitive, context-sensitive controls for personalization. By doing so, designers can help mitigate the inherent access friction that arises when a design optimized for one group excludes another.

% Moreover, the integration of features such as persistent user profiles and live previews not only streamlines the customization process but also helps maintain consistency across devices and applications. Ultimately, embracing these strategies can bridge the gap between the accessibility and visualization communities, fostering an ecosystem where the diverse needs of users are met without compromising the integrity and clarity of data representation.

Our vision of data visualization \textit{softerware} demands more involvement from research and industry. We want to encourage researchers to investigate further the needs of people with disabilities, designers to imagine new \textit{softerware} interfaces and interaction paradigms, and engineers to build robust systems that are capable of not only respecting a user's preferences and customizations, but providing persistence, interoperability, and system-class infrastructure.

\section{ACKNOWLEDGMENT}

We thank our ongoing conversations with MIT's Visualization Group as well as CU Boulder's Data and Design Group. This work was funded and made possible by Highsoft, with special thanks to Jørgen Tistel.

\pagebreak
\begin{IEEEbiography}{Frank Elavsky,}{\,}PhD student in the Human-Computer Interaction Institute at Carnegie Mellon University. His research interests include interactive data visualization, accessibility, and tool-making. His website is at \url{https://www.frank.computer/}. Contact him at fje@cmu.edu. %\vfill\pagebreak
\end{IEEEbiography}

\begin{IEEEbiography}{Marita Vindedal,}{\,}Accessibility Specialist and Developer with Highcharts, Vik I Sogn, Norway. Has a background in software engineering and web development, and has the recent years been working more in the field of accessibility, user experience and data visualization.
\end{IEEEbiography}

\begin{IEEEbiography}{Ted Gies} leads the Digital Accessibility Team at Elsevier and chairs the RELX Accessibility Guild. He has worked across RELX companies as a UX designer, researcher, and global accessibility authority. He chaired the company accessibility policies, developing several tools and coalitions across divisions. Ted has led design and user studies in Oil \& Gas Geoscience, Academic Research, and Corporate Innovation. He collaborates with many leading universities and IT vendors on accessibility. Gies has a patent award on a file naming system user interface and has published in several scholarly journals.
\end{IEEEbiography}

\begin{IEEEbiography}{Patrick Carrington,}{\,}Assistant Professor in the Human-Computer Interaction Institute at Carnegie Mellon University. His research emphasizes the design of systems to support people with diverse abilities. He studies mobile and wearable technology, builds assistive devices, and explores how computing can be used to support empowerment, independence, and improved quality of life. His current projects span topics from accessing digital content and media to developing technologies to support athletes with disabilities. Dr. Carrington has multiple conference and journal publications, winning Best Paper and Honorable Mention awards at the CHI and ASSETS conferences.
\end{IEEEbiography}

\begin{IEEEbiography}{Dominik Moritz,}{\,}Assistant Professor in the Human-Computer Interaction Institute at Carnegie Mellon University. His group’s research develops interactive systems that empower everyone to effectively analyze and communicate data. Dominik also manages the visualization team in Apple’s machine learning organization. His systems have won awards at academic venues (e.g. IEEE VIS and CHI), are widely used in industry, and by the Python and JavaScript data science communities. His website is at \url{https://www.domoritz.de/}.
\end{IEEEbiography}

\begin{IEEEbiography}{Øystein Moseng,}{\,}Chief Product Officer with Highsoft, Vik i Sogn, Norway. Moseng has a background in software engineering, and has held a specific focus in recent years on the intersection between data visualization, user experience, and accessibility. Contact him at oystein@highsoft.com.
\end{IEEEbiography}

\end{document}